\documentclass[12pt]{article}
\input{epsf}

\begin{document}

\begin{flushright}
\mbox{
\begin{tabular}{l}
    FERMILAB-PUB-99/032-T
\\ hep-ph/9903231
\end{tabular}}
\end{flushright}
\vskip 1.5cm
\begin{center}
\Large
{\bf Distinguishing $WH$ and $W b \bar{b}$ production \\
at the Fermilab Tevatron} \\
\vskip 0.7cm
\large
Stephen Parke and Sini\v{s}a Veseli \\
\vskip 0.1cm
{\small Theoretical Physics Department \\
Fermi National Accelerator Laboratory\\
P.O. Box 500, Batavia, IL 60510} \\
\vskip 0.5cm
March 2, 1999
\end{center}

\thispagestyle{empty}
\vskip 0.7cm

\begin{abstract}
The production of a Higgs boson in association with a $W$-boson
is the most likely process for the discovery of a light Higgs
at the Fermilab Tevatron. Since it decays primarily
to $b$-quark pairs, the principal background for
this associated Higgs production
process is $Wb\bar{b}$, where the $b\bar{b}$ pair comes from
the splitting of
an off mass shell gluon.
In this paper we investigate whether the spin angular correlations of
the final state particles can be used to
separate the Higgs signal from the $Wb\bar{b}$ background.
We develop a general numerical technique which allows one to
find a spin basis optimized according to a given criterion, and
also give a new algorithm for reconstructing the $W$
longitudinal momentum which is suitable for the
$WH$ and $Wb\bar{b}$ processes.
\end{abstract}
\newpage

\section{Introduction}
\label{intro}

At present, the existence of a neutral Higgs boson is certainly the
largest unresolved problem in the standard model (SM). Its mass
is {\it a priori} unknown, but direct searches and precision
electroweak measurements constrain it to be
$90< M_H < 280{\;\rm GeV}$ at 95\% confidence level
\cite{ICHEP98}. At the Tevatron collider there is a possibility
to search for the SM Higgs using the decay mode $H \rightarrow b\bar{b}$
\cite{SMW}, and
the  most promising process is the associated Higgs production
\begin{equation}
p \bar{p} \rightarrow W(\rightarrow e\nu) H(\rightarrow b \bar{b})\ .
\label{wh}
\end{equation}
The Fermilab search is extremely important, especially because the mass range
which can be covered at the Tevatron ($100<M_H<130{\;\rm GeV}$)
is also one of the most challenging
regions for the LHC to look for the SM Higgs  \cite{ATLAS}.
With sufficiently large data sample the Higgs signal could
be extracted from the background by analyzing the $b\bar{b}$
mass distribution.
However, given the fact that there are several large backgrounds to
process (\ref{wh}), any technique which can provide additional handles
on distinguishing the signal from the background would be useful.

In this paper we investigate the possibility
of using the spin angular correlations for separating the associated
Higgs production from its principal background
at the Tevatron, the $Wb\bar{b}$ process
\begin{equation}
p \bar{p} \rightarrow W(\rightarrow e\nu) g^*(\rightarrow b \bar{b})\ .
\label{wbb}
\end{equation}
In the case of
$e^+e^-\rightarrow ZH/ZZ$ in \cite{MP3} it was shown that
spin angular correlations can provide useful information if
good spin bases are chosen.
Since the $q\bar{q}\rightarrow WH/Wb\bar{b}$ processes have the
same spin structure,
it is natural for one to ask a question whether a similar analysis
would be useful for distinguishing (\ref{wh}) and (\ref{wbb})
at the Tevatron. However, due to the
hadronic collider environment, and also to the complexity of the
$Wb\bar{b}$ amplitudes, it is obvious that in this case a numerical
approach for finding the best spin basis is more appropriate than the
approach used in \cite{MP3}.
For that reason we develop here a new method which allows one
to find a spin basis optimized according to a given criterion.
This technique is completely general in the sense that it
can be used for optimizing  spin basis
regardless of which or how many
processes are being considered.
We apply our method to $WH$ and $Wb\bar{b}$ processes, and
suggest several possible strategies which could add
new information in an experimental analysis.
We also discuss one of the major uncertainties related
to our analysis, and that is the $W$ momentum reconstruction.
Our results indicate that the method which
has been used in the literature can distort
angular distributions considerably, and is therefore
inadequate for our purposes.  Because of that we propose a new $W$
reconstruction algorithm whose effects on angular distributions are
significantly less destructive.

The remainder of the paper is organized as follows: in Section
\ref{ang_corr} we give all relevant definitions, describe
numerical method and suggest possible strategies
for finding the optimal spin basis. In Section \ref{nres}
we present our results for angular distributions,
and show the effects which $W$
reconstruction algorithm has on those. Conclusions are contained
in Section \ref{conc}.

\section{Angular correlations}
\label{ang_corr}

\begin{figure}[p]
\epsfxsize = 5.0in
\centerline{\hspace*{+0.1cm}\vbox{\epsfbox{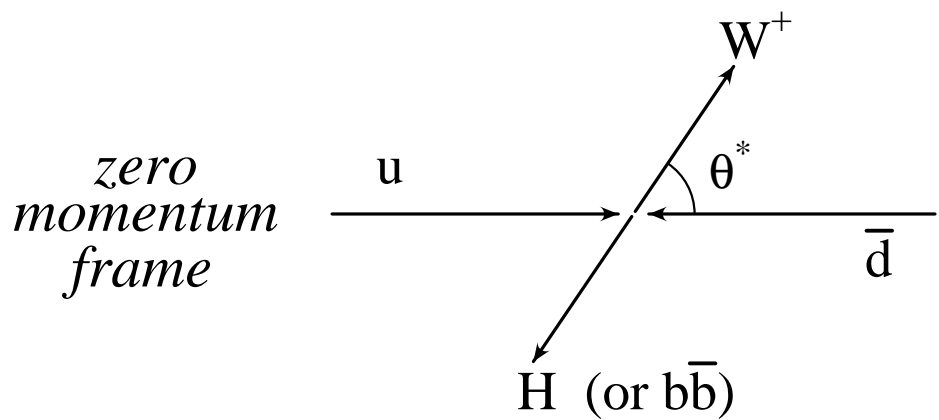}}}
\caption{Scattering angle $\theta^*$ in the zero momentum frame.
The up-quark comes from the proton beam more than 95\% of the
time at the Fermilab Tevatron.}
\label{zmf}
\end{figure}

\begin{figure}[p]
\epsfxsize = 5.0in
\centerline{\hspace*{+0.1cm}\vbox{\epsfbox{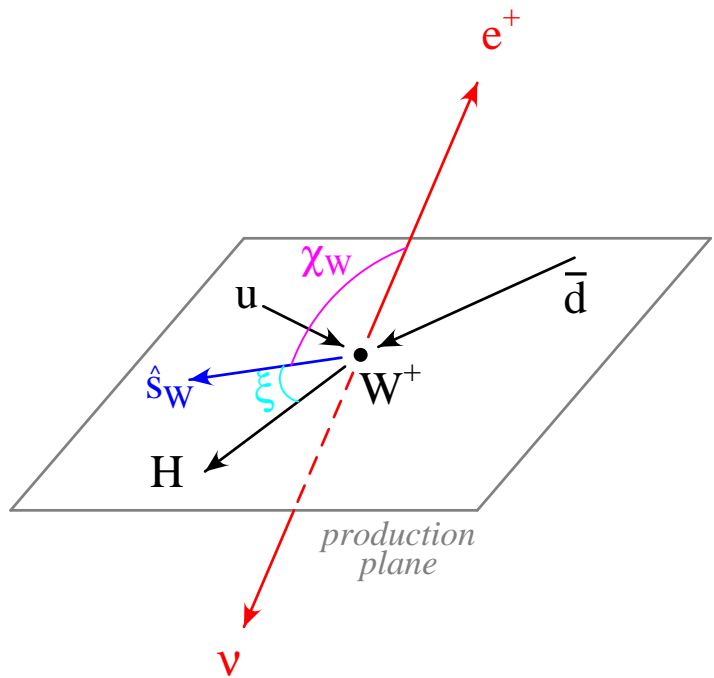}}}
\vspace*{-5.2cm}
\caption{Definitions for angles $\xi$ and $\chi_W$ in the $W$
rest frame.}
\label{chi_def}
\end{figure}

In order to apply the generalized spin-basis analysis \cite{PS} to processes
(\ref{wh}) and (\ref{wbb}) we first define the zero momentum frame (ZMF)
production angle $\theta^*$ ($0\leq \theta^* < \pi$) as the angle
between the incoming up-quark and
the $W$-boson produced in the
$q\bar{q}'\rightarrow W  X$ process (see Figure \ref{zmf}), where $X$ is
either $H$ or $g^*$.
The spin states for $W$ are defined in its
rest frame, where we decompose its spin along the vector $\hat{s}_W$,
which makes an angle $\xi$ with the $X$ particle
momentum in the clockwise direction.
The $X$ particle's spin can be decomposed in a similar way.
Relationship
between $\xi$ and $\theta^*$ determines specific spin basis in which
one can calculate  angular correlations among
the $WH$ and $Wg^*$ decay products in (\ref{wh}) and (\ref{wbb}).
These correlations involve distributions
of the angle $\chi_W$ ($\chi_{b\bar{b}}$)
that the charged lepton ($b$-quark) makes with the
spin vector of the $W$-boson ($b\bar{b}$ system).
Figure \ref{chi_def} illustrates the definitions for
angles $\xi$ and $\chi_W$.

In the case of $e^+e^-\rightarrow ZH/ZZ\rightarrow l\bar{l}\;jets$
the procedure for finding the optimal spin basis was based
on separating the polarized amplitudes for
$e^+e^-\rightarrow ZH/ZZ$ \cite{MP3}. In particular, it was shown that very
good separation between the $ZH$ and $ZZ$ events can be obtained
in the {\em transverse basis}, in which the longitudinal component of the
$ZH$ matrix element is zero by construction.

Since the amplitude for the process
$q\bar{q}'\rightarrow WH$ has the same spin structure as the one
for $e^+e^-\rightarrow ZH$,\footnote{The spin structure for the
process $u\bar{d}\rightarrow WH$ can be found in Eqs. (4)-(6)
in \cite{MP3}. The full matrix element squared, including the decay of
$W$-boson, can be obtained from Eq. (2) in \cite{KS}.}
the transverse basis is also a good starting
point for examining the $\cos{\chi}$ distributions in the $WH$ and $Wb\bar{b}$
processes. It is defined by
\begin{equation}
\tan{\xi} = {\tan{\theta^*}\over \sqrt{1-\beta^2_W}} \ ,
\end{equation}
where $\beta_W$ is the ZMF speed of the $W$-boson.
Nevertheless, due to the complex nature of the $q\bar{q}'\rightarrow Wg^*$
amplitudes, and also
to the fact that in $p\bar{p}$ collisions the center-of-mass energy
$\sqrt{\hat{s}}$ is not fixed, the approach of Ref. \cite{MP3}
for finding the optimal spin basis is not practical for our purposes here.
Because of that, instead of trying to separate polarized cross sections for
$q\bar{q}'\rightarrow WH/Wg^*$, we attempt to find the best basis
for processes (\ref{wh}) and (\ref{wbb})
by distinguishing the $\cos{\chi}$ distributions directly, using
a suitable multidimensional maximization procedure.

The basic idea of our method is to
divide $\cos{\theta^*}-\cos{\xi}$ plane into $n\times m$ regions, and
to associate with each of those a histogram containing distribution
in $\cos{\chi}$.\footnote{In general we do not know the up-quark
momentum direction. However, the up-quark comes from the proton beam
more than 95\% of the time at the Tevatron, and
therefore we will use proton direction
instead of the up-quark direction in defining $\cos{\theta^*}$
for the rest of this paper.}
A specific spin basis
is defined by choosing one of the $\cos{\xi}$ bins for all
of $n$ $\cos{\theta^*}$ bins, while the total $\cos{\chi}$ distribution
is obtained by summing contributions over the entire $\cos{\theta^*}$ range.
In other words, if $\cos{\xi_i}$ describes the spin vector of $W$
(or $b\bar{b}$ system) in the $i$-th
$\cos{\theta^*}$ bin, and $\sigma_i$ is the corresponding contribution
to the cross section, we have
\begin{equation}
\frac{d\sigma}{d\cos{\chi}}=\sum_{i=1}^n
\frac{d\sigma_i(\cos{\xi_i})}{d\cos{\chi}}\ .
\label{cchi_eq}
\end{equation}
In this way, by changing the $n$ $\cos{\xi_i}$ variables
using multidimensional maximization algorithm,
one can easily vary the definition of the spin basis
until the optimal separation
of the $WH$ and $Wb\bar{b}$ events is achieved.

Results of this procedure will clearly depend on which
criterion is used for determining the best possible separation of the
signal and the background.
We investigate here two possible criteria. The first one is based
on distinguishing between the shapes of the $\cos{\chi}$ distributions
for the two processes, and the function which we decided to maximize is given
by
\begin{equation}
\left|
\frac{1}{\sigma_{WH}}
\int d\cos{\chi}
\frac{d\sigma_{WH}}{d\cos{\chi}}\cos{\chi}
\right|
-
\left|
\frac{1}{\sigma_{Wb\bar{b}}}
\int d\cos{\chi}
\frac{d\sigma_{Wb\bar{b}}}{d\cos{\chi}}\cos{\chi}
\right|\ .
\label{fmin}
\end{equation}
With this criterion the resulting spin basis tends
to give $\cos{\chi}$ distributions
which are asymmetric for the $WH$ signal events
and symmetric for the $Wb\bar{b}$ background events.

The second criterion which we examine is based on maximizing significance
$S/\sqrt{B}$, where
$S$ and $B$ correspond to the number of events for the signal
and background, respectively:
\begin{eqnarray}
S&\propto&
\int_{\cos{\chi_{min}}}^{\cos{\chi_{max}}}  d\cos{\chi}
\frac{d\sigma_{WH}}{d\cos{\chi}}\ ,\\
B&\propto&
\int_{\cos{\chi_{min}}}^{\cos{\chi_{max}}}  d\cos{\chi}
\frac{d\sigma_{Wb\bar{b}}}{d\cos{\chi}} \ .
\end{eqnarray}
Once a particular spin basis is chosen and
the $\cos{\chi}$ distribution for both processes
is calculated using (\ref{cchi_eq}),
we choose angles ${\chi_{min}}$ and
${\chi_{\max}}$ in such a way to maximize the ratio $S/\sqrt{B}$.
Note that in the above coefficients of proportionality include the NLO
$K$-factors, our assumptions on the integrated luminosity and
double $b$-tagging efficiency, etc.

The main advantage of the method described above is that it offers
a systematic approach for investigating the possibility
of using the spin angular correlations to distinguish
signal events from the background, regardless of which or how many
processes are being considered.
For example, even though we are concerned here
only with the leading order
$Wb\bar{b}$ process as the most important background for the
associated Higgs production at the Tevatron, it would be straightforward
to include other backgrounds or the next-to-leading order effects
as well.
Note however that calculation of
angular correlations between the spin vector of
an intermediate gauge boson and momenta of its decay products
requires complete reconstruction of an event.
That is a major difficulty in the
case of the $WH/Wb\bar{b}$ production where longitudinal component
of neutrino is unknown. This issue will be discussed in more details
in the following section.

\section{Numerical results}
\label{nres}

Since the procedure outlined above
requires large statistics in order to make
errors in $d\sigma/d\cos{\chi}$ distributions as small as possible,
for the results presented in this paper we generated
about $10^8$ events (for each process), using the
{\em VEGAS} algorithm \cite{L}.\footnote{Because of the
large statistics and the large number of histograms
required by our method, Monte Carlo
simulations which would include all
other background processes, or take into account
next-to-leading order corrections, would have to be done
using a parallel event generator \cite{SV,K}.}
Calculations were done with $n=10$ bins along the
$\cos{\theta^*}$ axis, and $m=1000$ bins along the $\cos{\xi}$ axis, while
the search for the optimal basis
was performed using the
{\em downhill simplex method} \cite{NM,PTVF}.

Even though the analysis described in the previous section
can be performed for both $W$ and $b\bar{b}$ sides of an event,
we focus here only on the $\cos{\chi_W}$ distributions.
The reason is that
the correlations on the $W$ side of an event are much stronger
and provide us with more distinguishing power for separating
the $WH$ and $Wb\bar{b}$ processes.\footnote{On the  $b\bar{b}$ side
of the event we were unable to find a spin basis which would
considerably improve the small difference  between the
$WH$ and $Wb\bar{b}$ processes that was obtained
using the helicity basis.}

All results shown in this paper are obtained for the $W^+$ production
in $p\bar{p}$ collisions at $\sqrt{S}=2 {\;\rm TeV}$, with
the MRSR1 parton distribution functions ($\alpha_S(M_Z)=0.113$) \cite{MRS}.
In order to improve our lowest order cross sections, instead of natural
scales ($\mu\approx M_H$) we used somewhat lower scale of
$\mu=50{\; \rm GeV}$ \cite{EV}. At this scale the NLO $K$-factors
are about 1.1 for both $WH$ and $W b\bar{b}$ processes.
The Higgs mass was set to
$M_H=120{\; \rm GeV}$ and
the corresponding  $b\bar{b}$ mass range
to $102<M_{b\bar{b}}<141{\; \rm GeV}$.
In addition, we applied
the following set of isolation cuts and cuts
on rapidity and transverse momentum:
\begin{equation}
\begin{array}{rcl}
R_{b\bar{b}},R_{eb},R_{e\bar{b}}&>&0.7\ ,\\
|y_b|, |y_{\bar{b}}| &<& 2\ , \\
|y_e| &<& 2.5\ , \\
|p^T_b|, |p^T_{\bar{b}}| &>& 15{\;\rm GeV} \ ,\\
|p^T_e|, |p^T_{\nu}| &>& 20{\;\rm GeV} \ .
\end{array}
\label{cuts}
\end{equation}
Note the $|p^T_{\nu}|$ cut is the missing $E_T$ cut
and that the above cuts do not include
a cut on $\cos{\theta^*}$ \cite{KKY}.
Our results indicate that imposing the $\cos{\theta^*}$  cut actually
worsens our ability to separate the two processes based on the shape
of their $\cos{\chi_W}$ distributions, and therefore we did not
include it in the simulations based on
maximization of Eq. (\ref{fmin}). On the other hand, it is well known
that this cut can improve the $S/\sqrt{B}$ ratio by about 10\% \cite{KKY}.
Because of that, we take it into account for simulations based
on the significance criterion.

\begin{figure}[p]
\epsfysize = 5.0in
\centerline{\vbox{\epsfbox{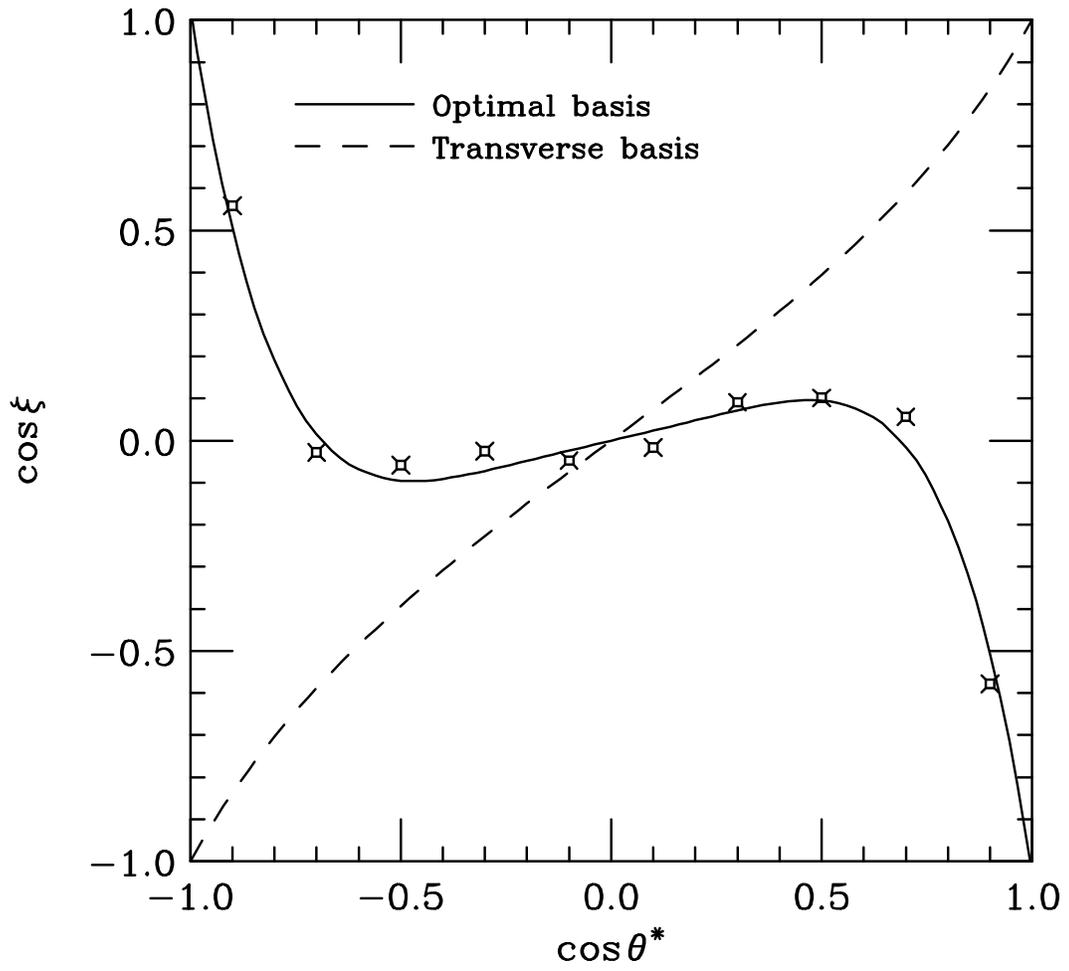}}}
\caption{The optimal basis for the shape criterion (points), together
with its $k=3$ polynomial approximation
(solid line) and with
the transverse basis for $\beta_W=0.67$ (dashed line).}
\label{basis}
\end{figure}

\begin{figure}[p]
\epsfysize = 5.0in
\centerline{\vbox{\epsfbox{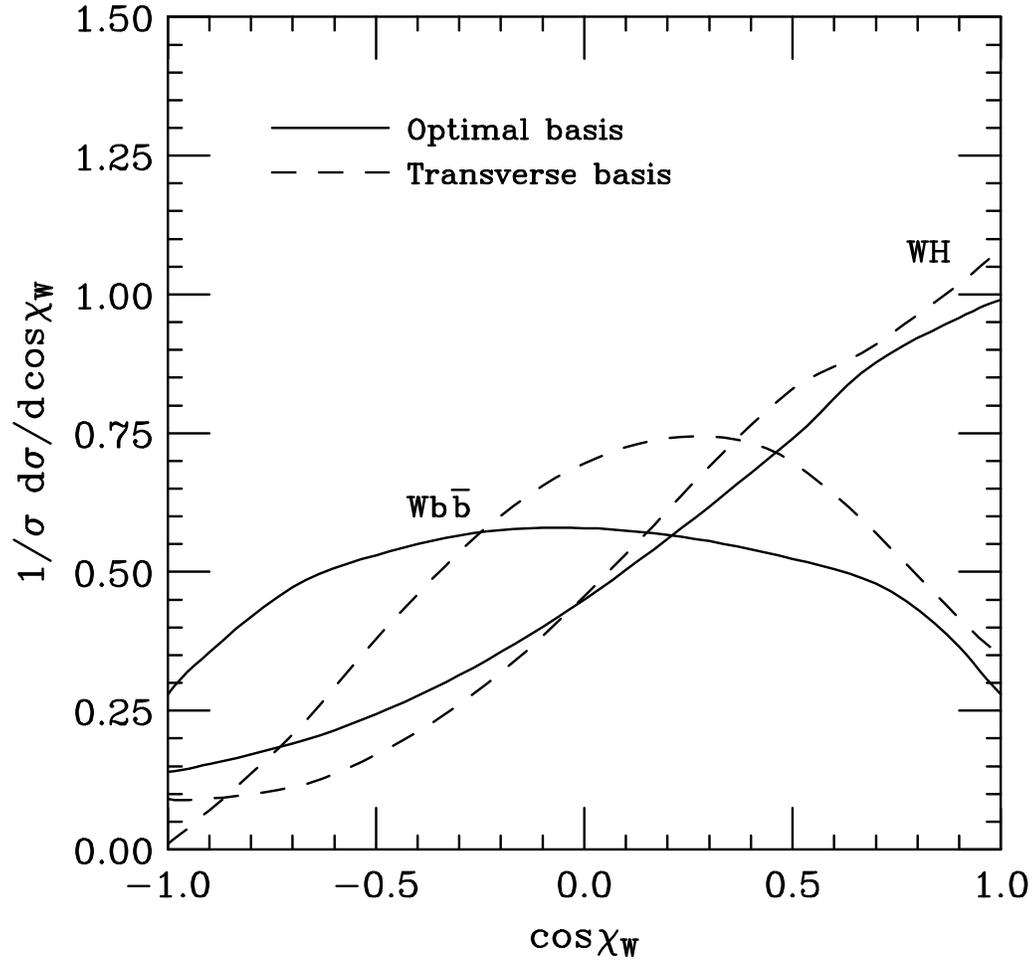}}}
\caption{Normalized $\cos{\chi_W}$ distributions for the polynomial
approximation of the optimal basis
(solid lines), and for the transverse basis (dashed lines).}
\label{cchi}
\end{figure}

\begin{figure}[p]
\epsfysize = 5.0in
\centerline{\vbox{\epsfbox{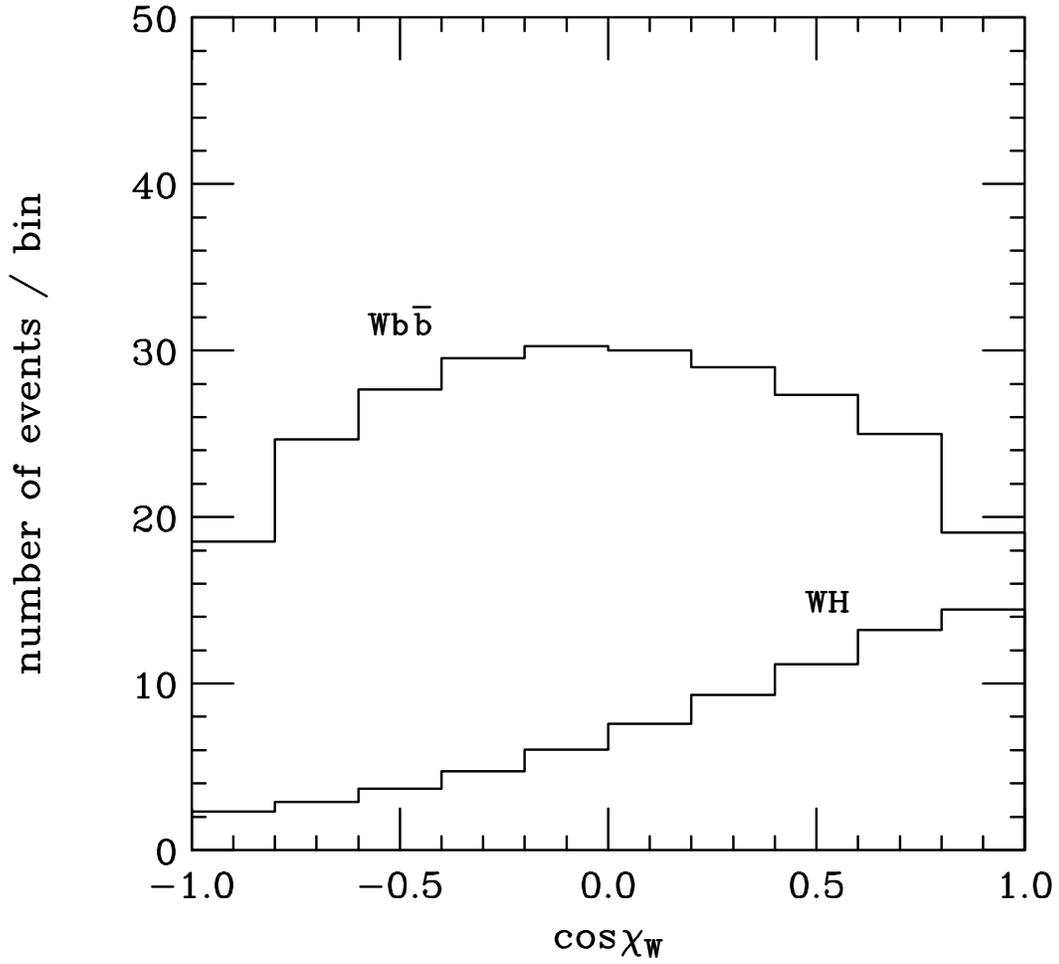}}}
\caption{Distribution of the number of events per bin in the polynomial
approximation of the optimal basis.
The total number of events for $WH$ is 75, and for $Wb\bar{b}$ is 260.}
\label{events_fb}
\end{figure}

\begin{figure}[p]
\epsfysize = 5.0in
\centerline{\vbox{\epsfbox{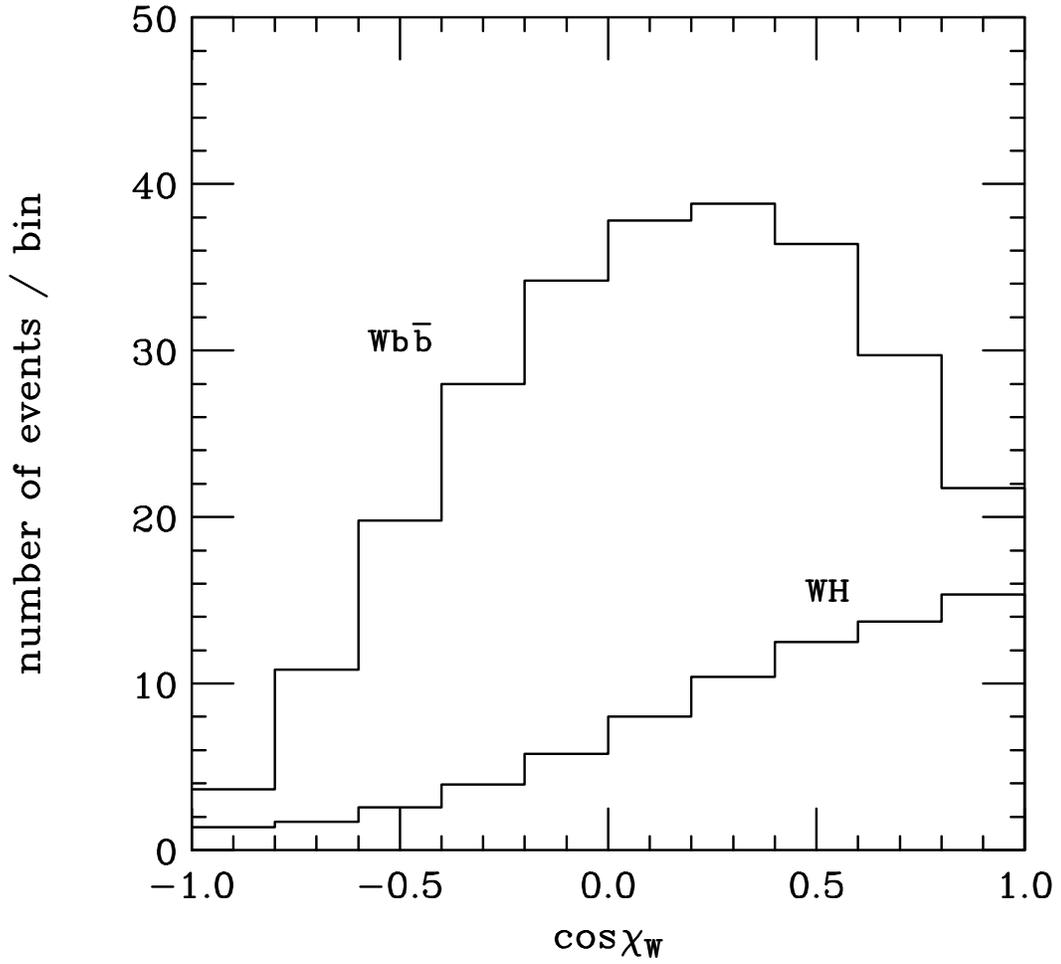}}}
\caption{Distribution of the number of events per bin in the transverse
basis.
The total number of events for $WH$ is 75, and for $Wb\bar{b}$ is 260.}
\label{events_tb}
\end{figure}

We first discuss our results obtained with the shape criterion.
In this case we found that the optimal basis
can be well approximated by the polynomial
of the form
\begin{equation}
\cos{\xi}=\sum_{i=1}^{k} a_i (\cos{\theta^*})^{2i-1}\ .
\label{fit1}
\end{equation}
In particular, in Figure \ref{basis} we compare the exact result
obtained by maximization procedure to polynomial with $k=3$ and
coefficients
\begin{eqnarray}
a_1 &=& 0.2354\ , \nonumber \\
a_2 &=& 0.1808\ ,
\label{fit2}\\
a_3 &=& -1.442\ . \nonumber
\end{eqnarray}
There we also show the
transverse basis  with specific choice of $\beta_W=0.67$, which
is close to
the averages of the $\beta_W$ distributions for both
$WH$ and $Wb\bar{b}$ processes (0.68 and 0.66, respectively).
The actual normalized $\cos{\chi_W}$ distributions corresponding
to the polynomial
approximation of the  optimal basis and for the
transverse basis are given in Figure \ref{cchi}. As expected,
in the optimal basis
the $Wb\bar{b}$ distribution is nearly symmetric.
Figures \ref{events_fb} and \ref{events_tb} illustrate what one might
expect in terms of the number of events per bin in those two bases.
These results were obtained by multiplying our $W^+$ cross sections
by four to take into account contributions from the $W^-$ production and
the contribution from the $W^{\pm}$ decays into muons, by taking
into account the NLO $K$-factor of 1.1 for both $WH$ and
$Wb\bar{b}$ processes,
and also by assuming the double $b$-tagging efficiency of $\epsilon_b^2=0.45$
and integrated luminosity of $10{\rm\; fb}^{-1}$.
We would like to point out here that the shape of the $Wb\bar{b}$
$\cos{\chi_W}$ distribution is significantly different in the two bases
being discussed. On the other hand, this is not the case
for the $WH$ process. Clearly, the difference in the shape
of the $\cos{\chi_W}$ distributions under the change of spin basis
may provide an additional handle for separating the two processes.

\begin{figure}[p]
\epsfysize = 5.0in
\centerline{\vbox{\epsfbox{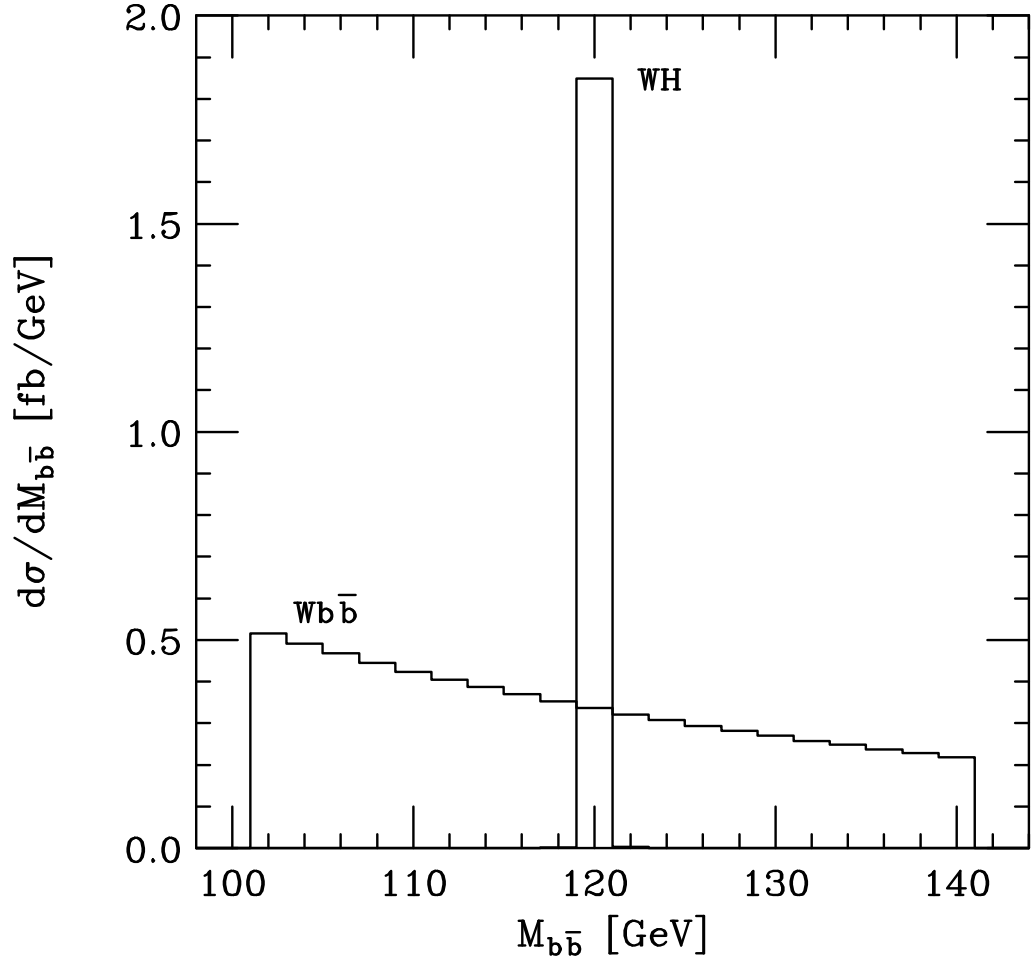}}}
\caption{$M_{b\bar{b}}$ distribution of the cross section.
No smearing of the $b$-quark jet energies has been performed here.}
\label{mbb}
\end{figure}

\begin{figure}[p]
\epsfysize = 5.0in
\centerline{\vbox{\epsfbox{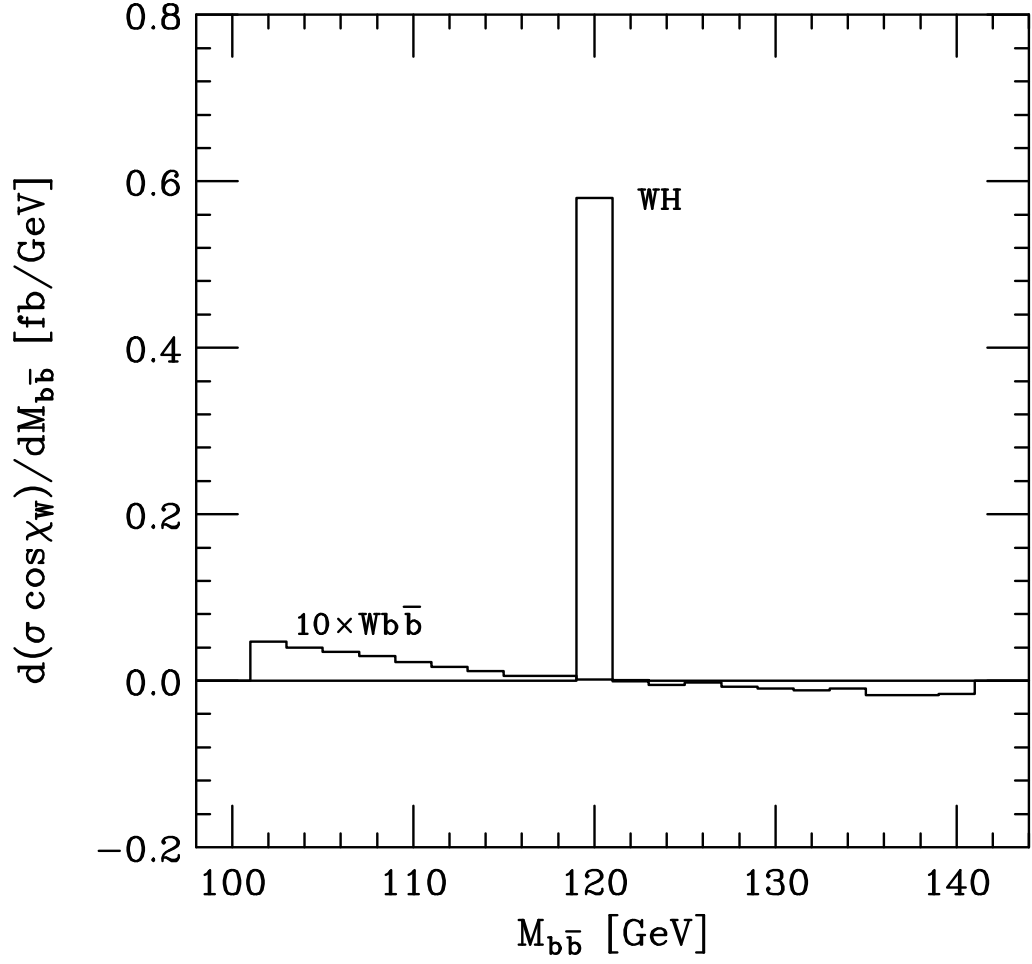}}}
\caption{$M_{b\bar{b}}$ distribution of $\sigma \cos{\chi_W}$
in the polynomial approximation of the optimal basis.}
\label{mbbcchi}
\end{figure}

\begin{figure}[p]
\epsfysize = 5.0in
\centerline{\vbox{\epsfbox{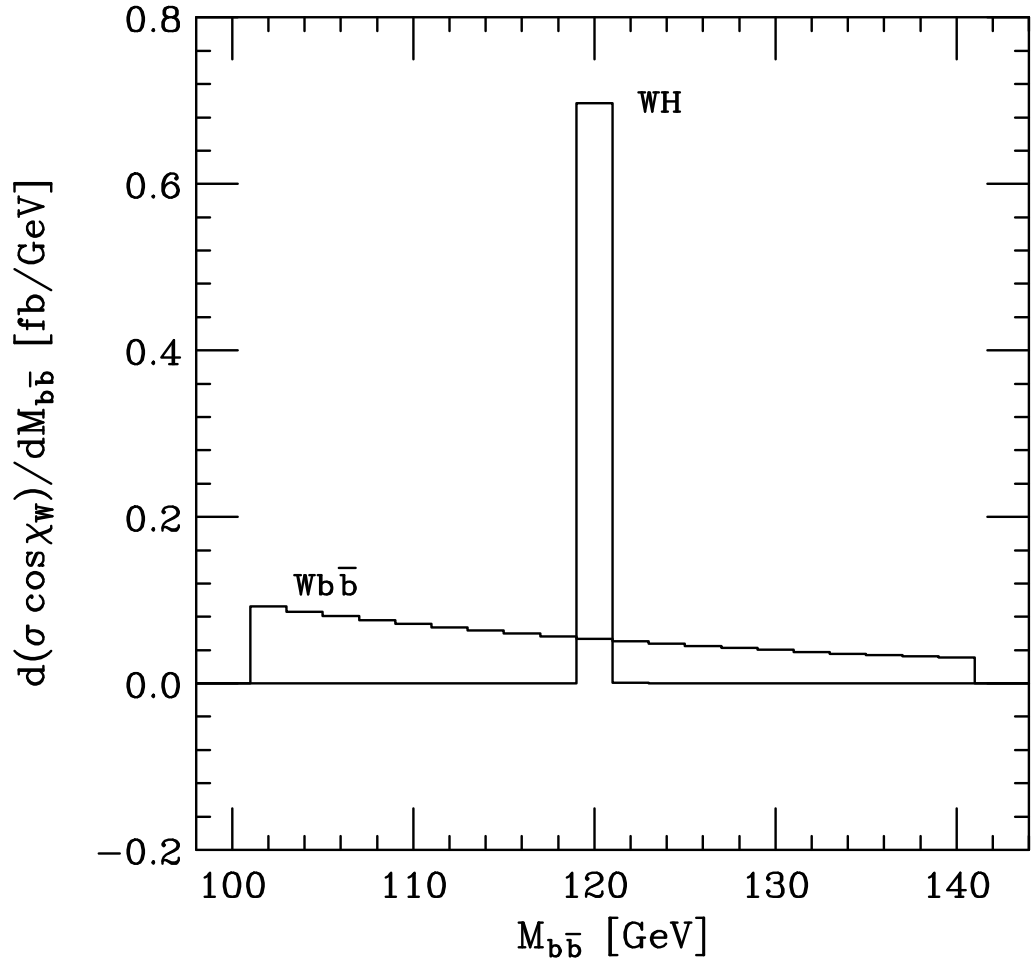}}}
\caption{$M_{b\bar{b}}$ distribution of $\sigma \cos{\chi_W}$
in the transverse basis.}
\label{mbbcchi_tb}
\end{figure}

Another interesting possibility of using angular correlations
for distinguishing between the signal and the background is illustrated
in Figures \ref{mbb} through \ref{mbbcchi_tb}. Instead of looking
at $d\sigma/d\cos{\chi_W}$ directly, we investigate the
$M_{b\bar{b}}$ distributions of the quantity $\sigma\cos{\chi_W}$.
Those distributions vanish in
the spin basis in which $d\sigma/d\cos{\chi_W}$ is perfectly
symmetric, because
in evaluating $d(\sigma\cos{\chi_W})/dM_{b\bar{b}}$ one
effectively integrates over $\cos{\chi_W}$. This is precisely
the reason why our optimal basis reduces the background in
$\sigma\cos{\chi_W}$ much more efficiently than the transverse
basis, as can be seen by comparing Figures \ref{mbbcchi} and \ref{mbbcchi_tb}.
However, at this point one should also observe that the main disadvantage of
analyzing quantities such as
$\sigma\cos{\chi_W}$
is the inclusion of the statistical errors of the entire
$\cos{\chi_W}$ distribution, which may limit its potential usefulness
in an experimental analysis with small statistics.

In our simulations based on maximizing significance
the number of events for both
signal and background was obtained by summing all decay channels, and
under the same assumptions as before ($K$-factors of 1.1,
$\epsilon_b^2=0.45$, and $\int {\cal L}\;dt= 10{\rm\; fb}^{-1}$).
As mentioned earlier, besides cuts given in (\ref{cuts}), here we also take
into account  a cut on $\cos{\theta^*}$ \cite{KKY},
\begin{equation}
|\cos{\theta^*}|\leq \cos{\theta^*_{max}}\ ,
\end{equation}
which can increase
the ratio $S/\sqrt{B}$ by about 10\% (see Table \ref{tbl}).
Using the method described in Section \ref{ang_corr}, and for any given value
of $\cos{\theta^*_{max}}$, we were able to find a spin basis
(and a set of cuts on $\cos{\chi}$) in which
one could further improve this ratio by additional 2-3\%.

\begin{figure}[p]
\epsfysize = 5.0in
\centerline{\vbox{\epsfbox{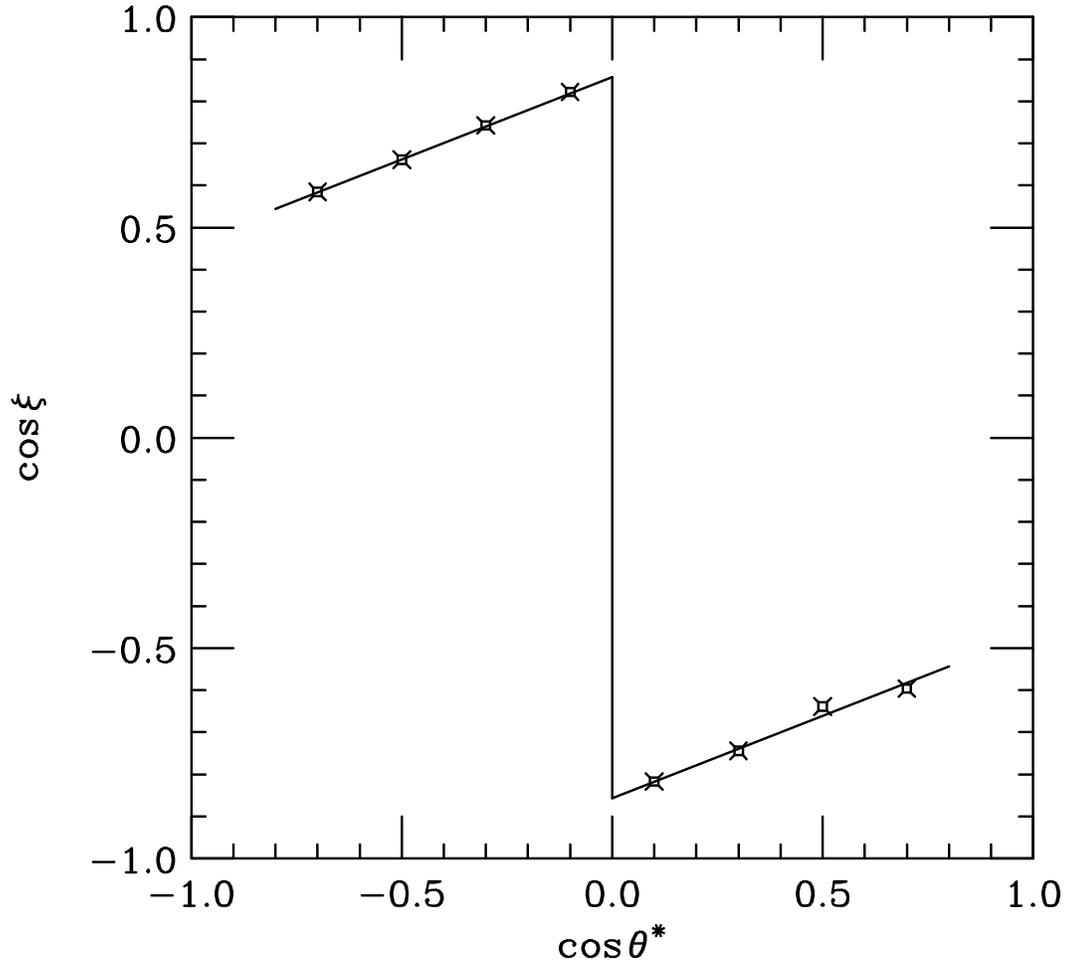}}}
\caption{The optimal basis for the significance criterion with
$\cos{\theta_{max}}=0.8$ (points), together with its
approximation given by Eq. (\protect\ref{lfit}) (solid line).
}
\label{sig_basis}
\end{figure}

\begin{figure}[p]
\epsfysize = 5.0in
\centerline{\vbox{\epsfbox{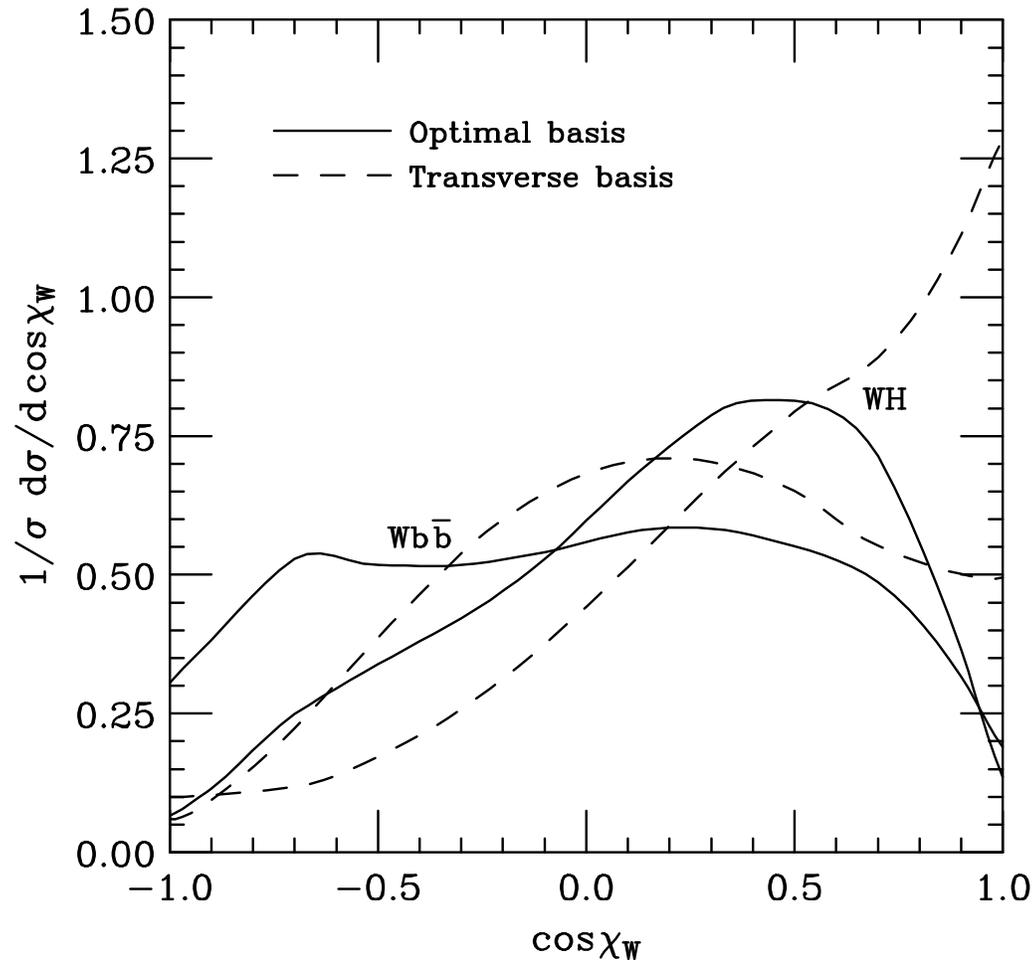}}}
\caption{Normalized $\cos{\chi_W}$ distributions for
the basis optimized according to the significance criterion with
$\cos{\theta_{max}}=0.8$
(solid lines), and corresponding results
for the transverse basis (dashed lines).}
\label{sig_cchi}
\end{figure}

\begin{figure}[p]
\epsfysize = 5.0in
\centerline{\vbox{\epsfbox{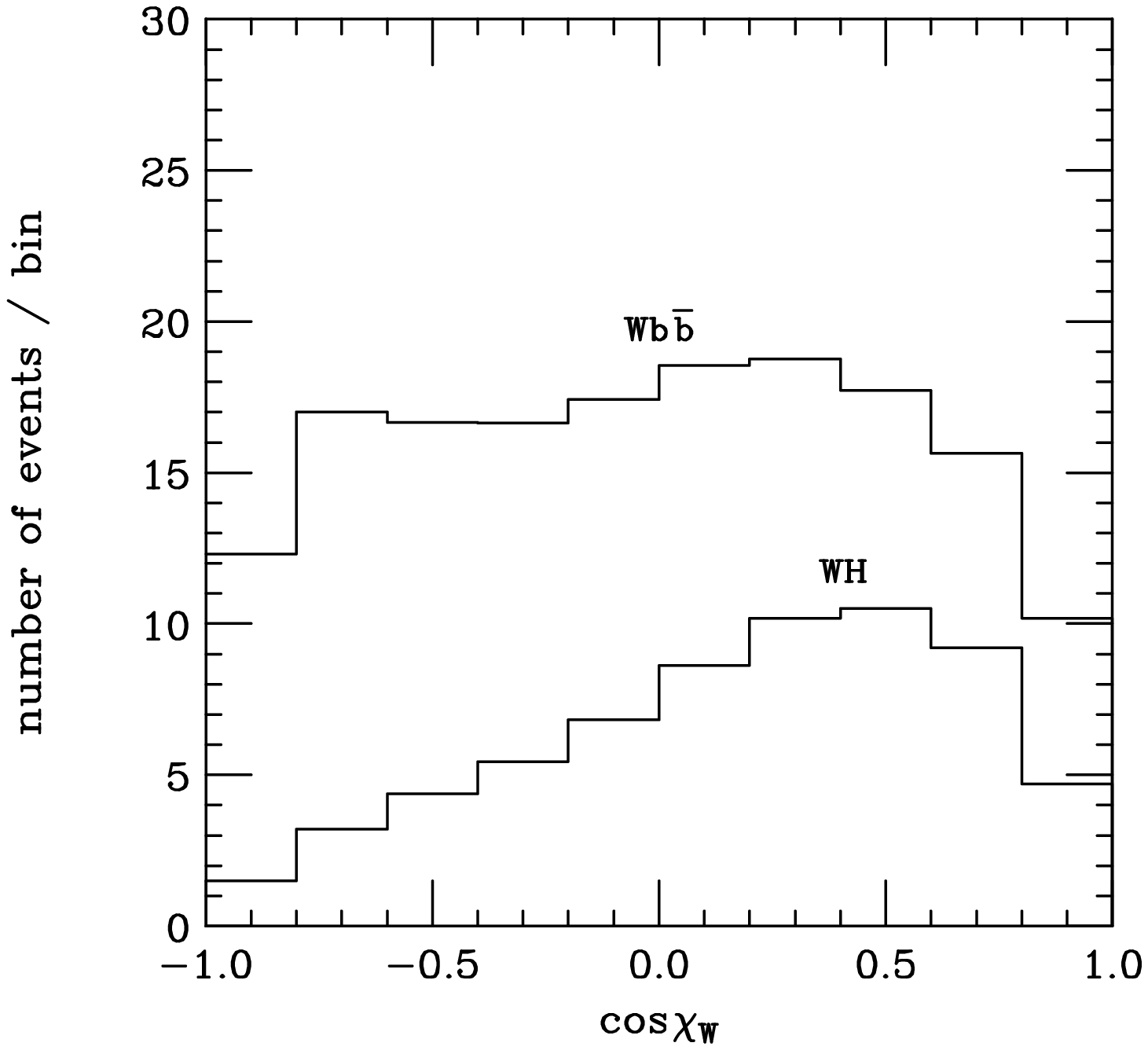}}}
\caption{Distribution of the number of events per bin in
the optimal basis
($\cos{\theta_{max}}=0.8$).  Without cuts on $\cos{\chi_W}$
we have 65 events for $WH$, and 161 events
for $Wb\bar{b}$ ($S/\sqrt{B}=5.12$). With cuts
$-0.6<\cos{\chi_W}<1.0$ these numbers are reduced to
60 and 131, respectively ($S/\sqrt{B}=5.24$).}
\label{sig_events_fb}
\end{figure}

\begin{figure}[p]
\epsfysize = 5.0in
\centerline{\vbox{\epsfbox{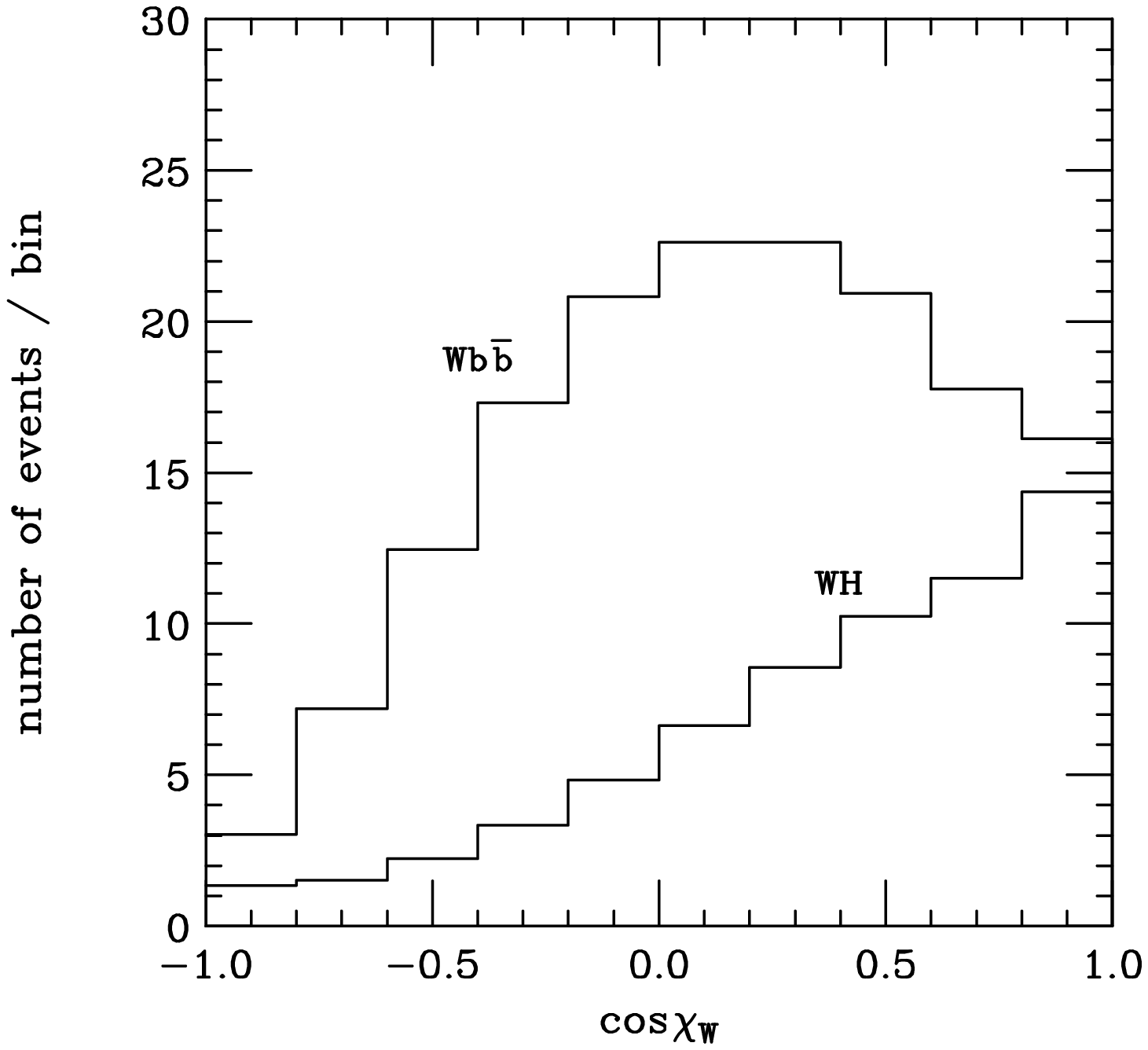}}}
\caption{Distribution of the number of events per bin in the transverse
basis
($\cos{\theta_{max}}=0.8$).
Without cuts on $\cos{\chi_W}$
we have 65 events for $WH$, and 161 events
for $Wb\bar{b}$ ($S/\sqrt{B}=5.12$). With cuts
$-0.1<\cos{\chi_W}<1.0$ these numbers are reduced to
54 and 110, respectively ($S/\sqrt{B}=5.15$).}
\label{sig_events_tb}
\end{figure}

Our results with $\cos{\theta_{max}}=0.8$ are shown in Figures
\ref{sig_basis} through \ref{sig_events_tb}.
Figure  \ref{sig_basis} shows the optimal basis definition. In this case
we found that it can be approximated by
\begin{equation}
\cos{\xi} = -0.857\;{\rm sign}(\cos{\theta^*})+0.391 \cos{\theta^*}\ .
\label{lfit}
\end{equation}
Normalized $\cos{\chi_W}$ distributions corresponding to the
optimal basis are compared
in Figure \ref{sig_cchi} to the results obtained using the transverse
basis.
Figures \ref{sig_events_fb} and \ref{sig_events_tb} illustrate what can be
expected in terms of the number of events per bin in those two bases.

\begin{figure}[tb]
\epsfysize = 5.0in
\centerline{\vbox{\epsfbox{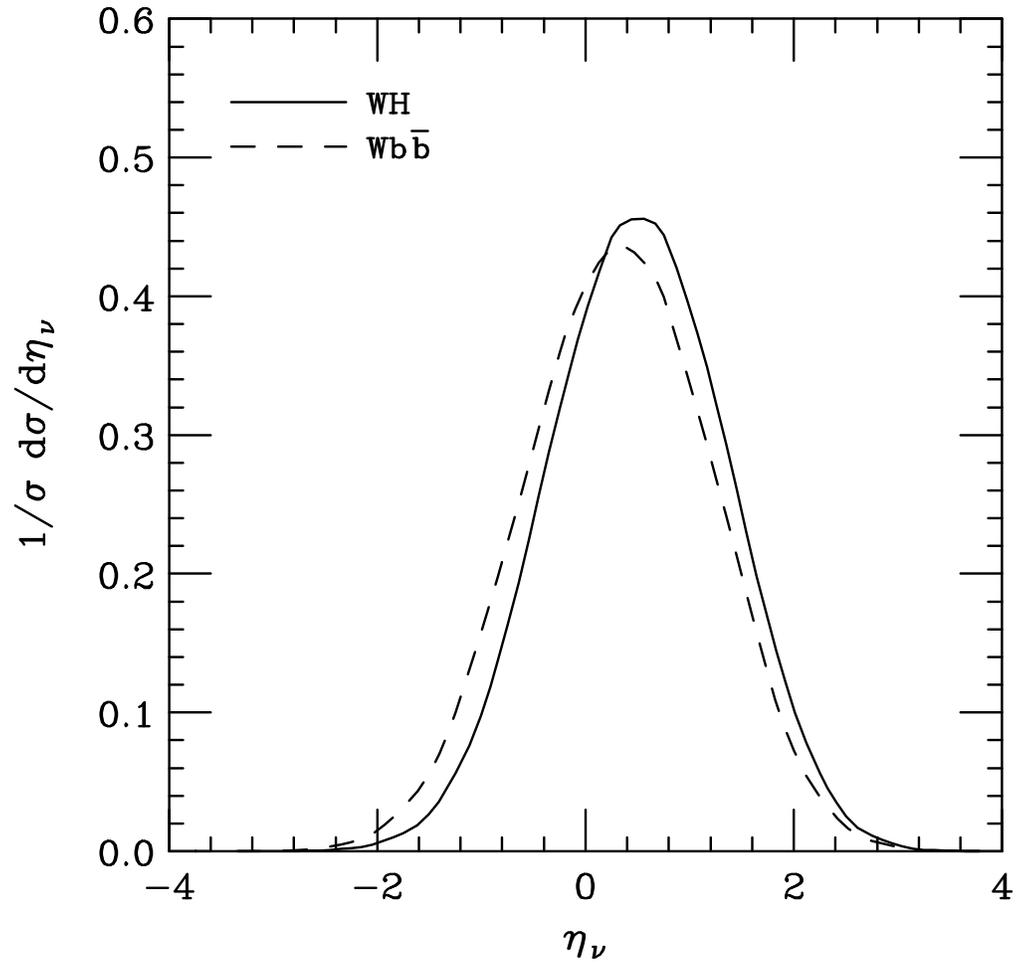}}}
\caption{Normalized $\eta_{\nu}$ distribution of the cross section
($W^+$ production).}
\label{ynu}
\end{figure}

\begin{figure}[tb]
\epsfysize = 5.0in
\centerline{\vbox{\epsfbox{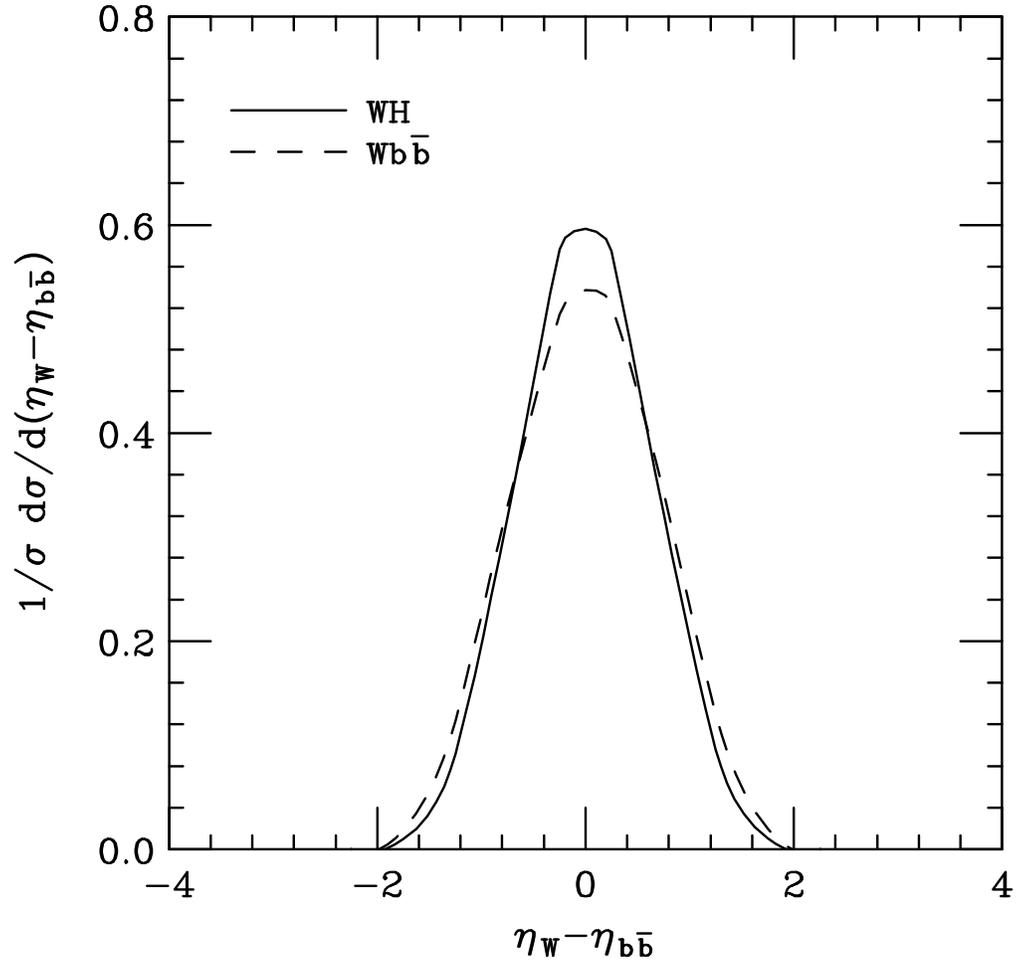}}}
\caption{Normalized $\eta_{W}-\eta_{b\bar{b}}$ distribution of
the cross section
($W^+$ production).}
\label{ywbb}
\end{figure}

Due to the fact that the longitudinal momentum of the
neutrino is unknown,
reconstruction of an event involving $W$-boson is the most important
problem related to the calculation of the spin
angular correlations which we discussed in this paper.
By assuming that $W$ is
on shell, and using $p_e$ and $p^T_{\nu}$ which are actually
measured, this component can be
reconstructed up to a two-fold ambiguity for a solution of a
quadratic equation.
The algorithm for choosing the correct solution
which has been used in the literature \cite{KKY} is based
on the asymmetry
of the neutrino rapidity distribution.
{}From  Figure \ref{ynu} it can be readily seen that
by choosing the larger (smaller)
solution for $p^z_{\nu}$ in the case of $W^+$ ($W^-$),
one can improve the probability
of finding the correct $W$ momentum.
Nevertheless, we propose here that for the $WH$ and $Wb\bar{b}$ processes
reconstruction algorithm
is based on the distribution of the difference between
the $W$ rapidity and the rapidity of the $b\bar{b}$ system
(see Figure \ref{ywbb}).
Since this distribution is peaked at zero,
our prescription consists of choosing the solution for
$p^z_{\nu}$ which results in a smaller absolute value for
$\eta_W-\eta_{b\bar{b}}$.
The advantage of using
$\eta_W-\eta_{b\bar{b}}$ instead of $\eta_{\nu}$
is that its distribution is narrower.
Furthermore, unlike the
$\eta_{\nu}$ distribution, it
is almost identical for $WH$ and $Wb\bar{b}$, which
means that our algorithm will work equally well for both processes.

\begin{figure}[tb]
\epsfysize = 5.0in
\centerline{\vbox{\epsfbox{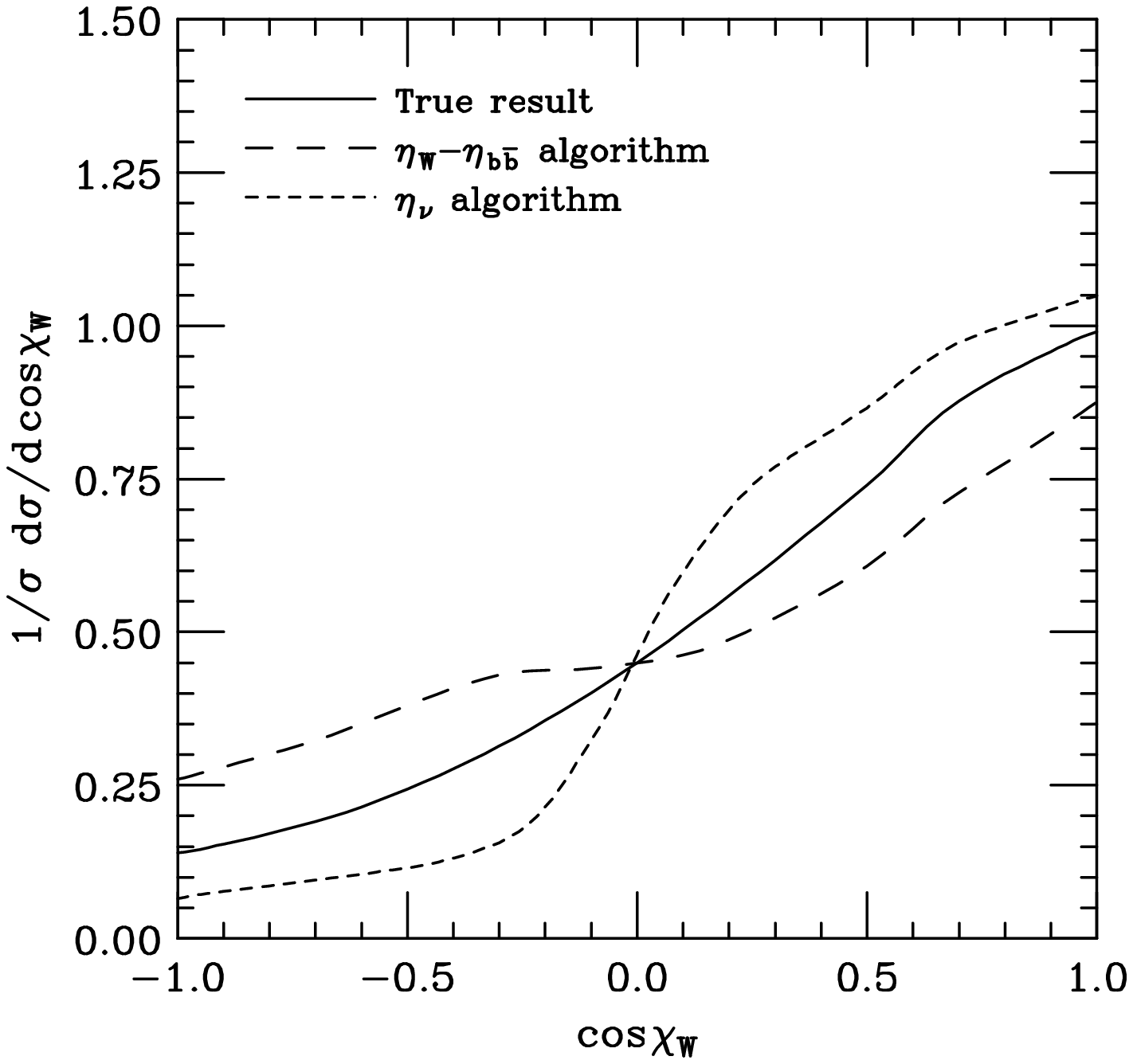}}}
\caption{Normalized $WH$ $\cos{\chi_W}$ distribution for the basis
optimized according to the shape criterion.
The true result is shown with
the full lines, while results obtained using
the $\eta_{W}-\eta_{b\bar{b}}$ and $\eta_{\nu}$ $W$ reconstruction
algorithms are plotted with the dashed lines and
short dashed lines, respectively.}
\label{cchi_fb_wh}
\end{figure}

\begin{figure}[tb]
\epsfysize = 5.0in
\centerline{\vbox{\epsfbox{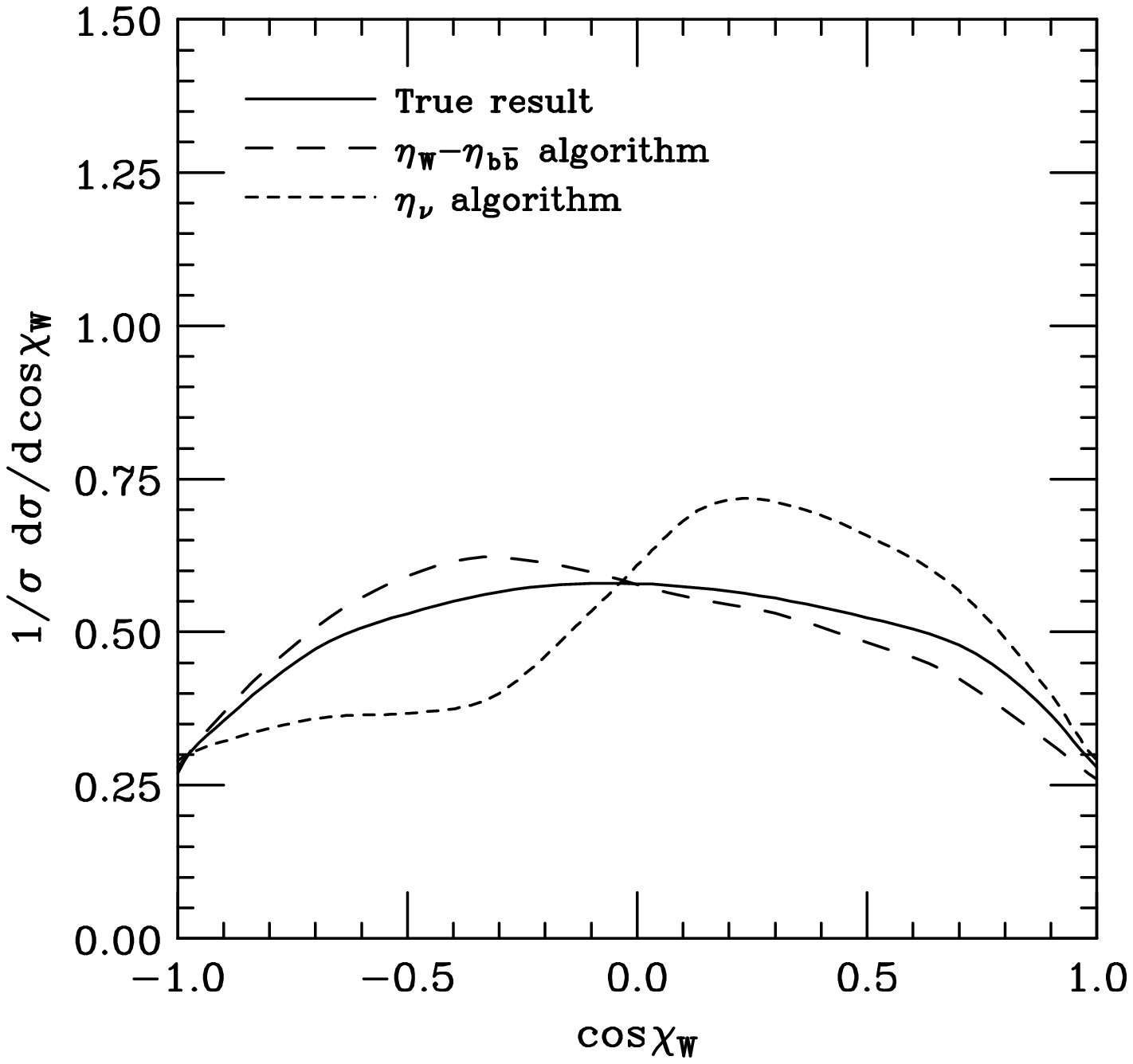}}}
\caption{Normalized $Wb\bar{b}$ $\cos{\chi_W}$ distribution
for the basis optimized according to the shape criterion.
The true result is shown with
the full lines, while results obtained using
the $\eta_{W}-\eta_{b\bar{b}}$ and $\eta_{\nu}$ $W$ reconstruction
algorithms are plotted with the dashed lines and
short dashed lines, respectively.}
\label{cchi_fb_wbb}
\end{figure}

\begin{figure}[tb]
\epsfysize = 5.0in
\centerline{\vbox{\epsfbox{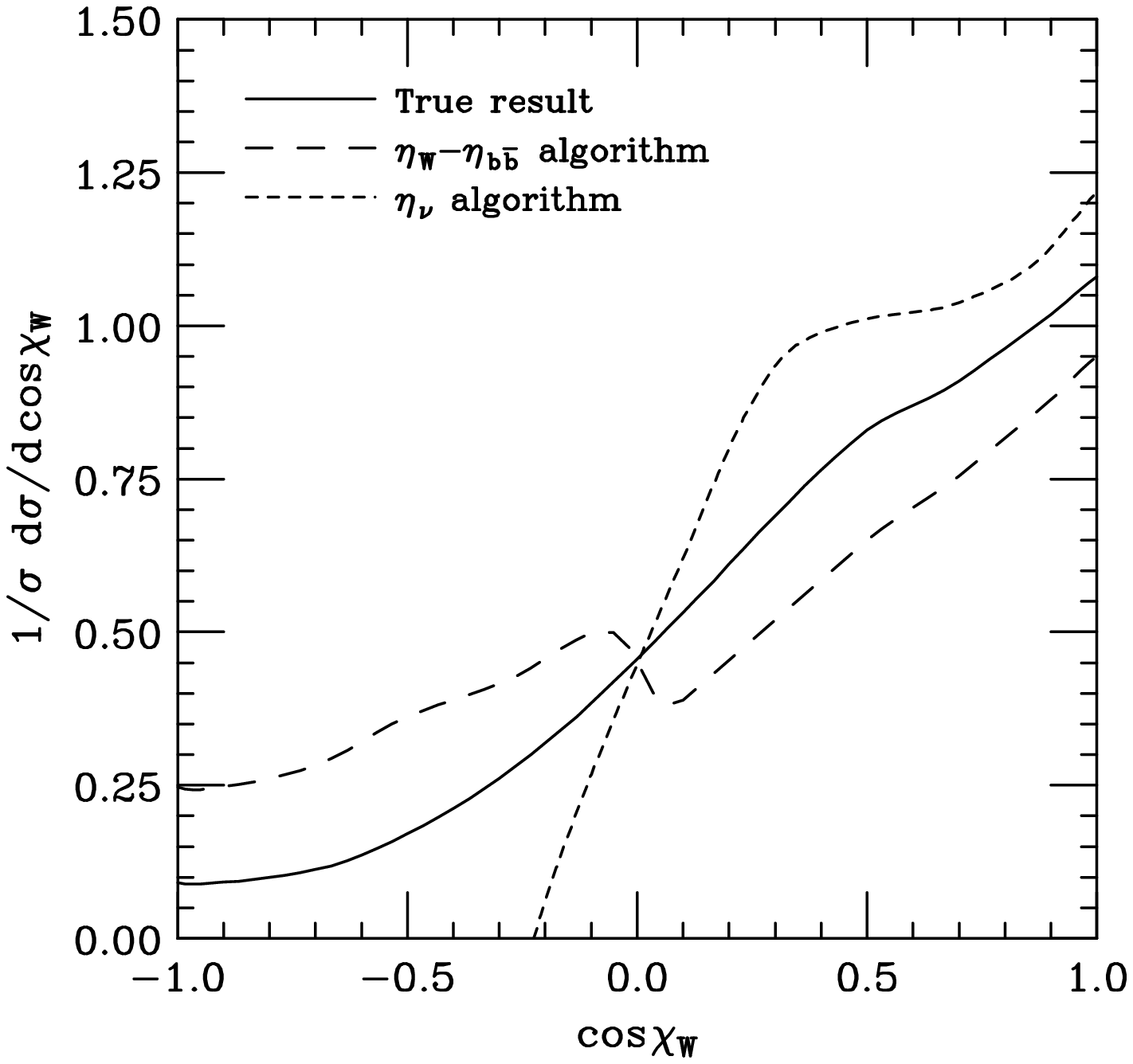}}}
\caption{Normalized $WH$ $\cos{\chi_W}$ distribution for the
transverse basis.
The true result is shown with
the full lines, while results obtained using
the $\eta_{W}-\eta_{b\bar{b}}$ and $\eta_{\nu}$ $W$ reconstruction
algorithms are plotted with the dashed lines and
short dashed lines, respectively.}
\label{cchi_tb_wh}
\end{figure}

\begin{figure}[tb]
\epsfysize = 5.0in
\centerline{\vbox{\epsfbox{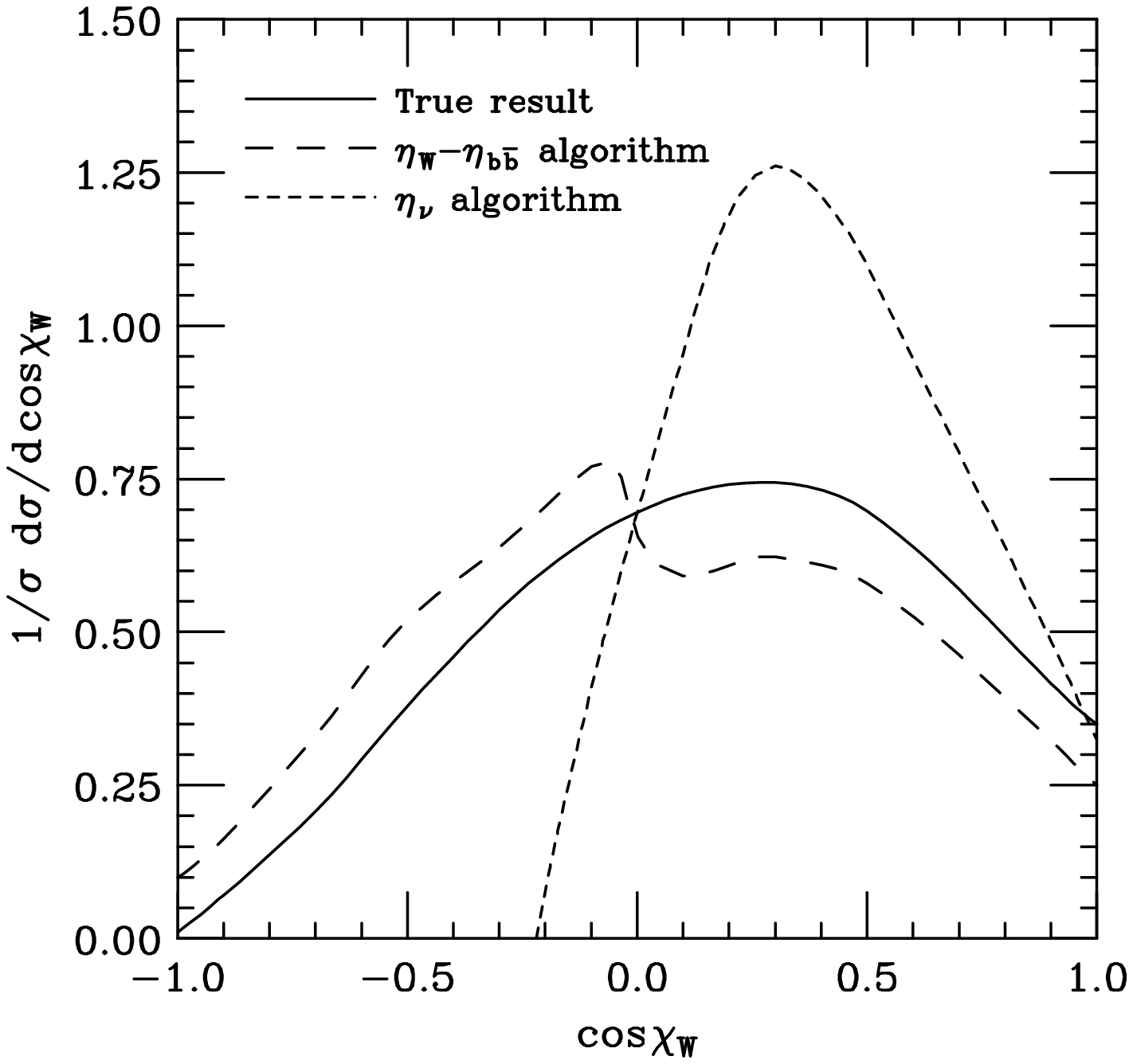}}}
\caption{Normalized $Wb\bar{b}$ $\cos{\chi_W}$ distribution
for the transverse basis.
The true result is shown with
the full lines, while results obtained using
the $\eta_{W}-\eta_{b\bar{b}}$ and $\eta_{\nu}$ $W$ reconstruction
algorithms are plotted with the dashed lines and
short dashed lines, respectively.}
\label{cchi_tb_wbb}
\end{figure}

In order to investigate the effect that $W$ reconstruction algorithm has
on $\cos{\chi_W}$ distributions, we have repeated calculations
shown on Figure \ref{cchi} (without
cuts on $\cos{\theta^*}$) for the polynomial approximation of
the optimal basis (Eqs. (\ref{fit1}) and (\ref{fit2})),
and for the transverse basis. Results given in Figures \ref{cchi_fb_wh}
through \ref{cchi_tb_wbb} show that the $\cos{\chi_W}$
distributions obtained using
our prescription for
reconstructing the $W$ momentum are much more closer to the exact curves
than are the ones obtained using the $\eta_{\nu}$ algorithm.
Note that
one of the reasons for distortion of the
reconstructed $\cos{\chi_W}$ distributions is the fact that
in our calculations the $W$ width is taken into account,
while the reconstruction algorithms assume the on-shell $W$.

Besides the issues related to reconstruction of the $W$ momentum,
another problem which might affect experimental analysis of the $\cos{\chi_W}$
distributions is the mismeasurement of the $b$-quark momenta.
We have simulated that by imposing
a Gaussian distribution of relative errors (with the variance of 5\%)
on both $b$ and $\bar{b}$ momenta, and our results indicate that these effects
are small.

\section{Conclusions}
\label{conc}

In this paper we investigated the possibility of
using the spin angular correlations for distinguishing between the $WH$
and $Wb\bar{b}$ processes at the Fermilab Tevatron.
We developed a general numerical method for finding the
spin basis optimized according to a given criterion, and
also suggested several possible strategies for utilizing this technique
in the Higgs search at the Tevatron.

Our simulations
indicate that the spin angular correlations may provide additional
handle on separating the signal from the background. Still, there
are several problems that would have to be solved for a
successful experimental analysis, and the largest one is certainly
the event reconstruction.
In this regard we  proposed a new $W$ reconstruction algorithm which
significantly reduces effects related to the $W$ momentum ambiguities.
We hope that this algorithm can be further improved upon.

The obvious extension of this work would involve including the NLO
corrections, as well including  the other background processes.
However, these calculations would be numerically quite challenging,
and before they are attempted a feasibility study of their usefulness
should be completed.

\section*{Acknowledgements}
We would like to acknowledge useful discussions with
M. Bishai, I. Dunietz, E. Eichten, R.K. Ellis, J. Goldstein, C. Hill,
J. Incandela, G. Mahlon, T.K. Nelson, F.D. Snider, L. Spiegel, and
D. Stuart.
Fermilab is operated by URA under DOE contract DE-AC02-76CH03000.

\newpage

\begin{table}
\caption{Expected number of events at
$10{\rm\; fb}^{-1}$
for the signal ($WH$)  and the background ($Wb\bar{b}$)
as a function of the $\cos{\theta^*}$ cut. Results shown
are obtained for a 120 GeV Higgs.
}
\begin{center}
\begin{tabular}{|c|ccc|}
\hline
$\cos{\theta^*_{\max}}$ & $S$ & $B$ & $S/\sqrt{B}$ \\
\hline
1.0 & 75 & 260 & 4.65 \\
0.9 & 70 & 198 & 4.97 \\
0.8 & 65 & 161 & 5.12 \\
0.7 & 58 & 131 & 5.07 \\
0.6 & 51 & 107 & 4.93 \\
0.5 & 44 & 87 & 4.71 \\
\hline
\end{tabular}
\label{tbl}
\end{center}
\end{table}

\clearpage
\newpage

\clearpage

\end{document}